# A Unified Electrostatic-to-Spin Framework for Asymmetric Multi-Gate CMOS Quantum Devices

Zeheng Wang[1,2,*], Yan Liu[1,2], Yue Hao[1,2], and Genquan Han[1,2]

[1] Hangzhou Institute of Technology, Xidian University, Hangzhou 311231, China

[2] Faculty of Integrated Circuit, Xidian University, Xi'an 710071, China

Email: zenwang@outlook.com

## Abstract

In advanced complementary metal-oxide-semiconductor (CMOS) quantum chips, compact gate stacks make it difficult to connect lithographic geometry, electrostatic confinement and many-electron spin filling in one transparent model. This connection is central to design-technology co-optimization (DTCO). Here we develop a reduced-order analytical framework for asymmetric multigate silicon quantum-dot devices. Its electrostatic core, the Poisson-kernel coupled-interface Green-function (PK-GF) model, agrees with an independent finite-volume solution at the millivolt scale for the matched two-dimensional problem, without fitting to that solution. We then pass the gate-derived confinement, rather than a harmonic or fitted potential, to a spin-valley many-body calculation for a jellybean quantum dot with N = 2-17 electrons at B = 5 T. The unrestricted Hartree-Fock (UHF) solution supports occupation-dependent, Wigner-molecule-like charge localization but likely overestimates spin polarization. Complete active-space configuration interaction (CASCI) supports a low-spin branch within the tested active spaces, which aligns with the experiments. The workflow therefore connects CMOS layout, device electrostatics, and potential-determined quantum observables, providing an auditable modelling layer for CMOS-based qubit design and DTCO.

## Introduction

Silicon quantum dots combine long-lived spin states with fabrication processes inherited from complementary metal–oxide–semiconductor (CMOS) technology, offering a credible route toward densely integrated quantum processors.[1–7] In these devices, nanoscale gate electrode does more than define a potential well: It determines tunnel barriers, orbital structure, electron occupancy and spin filling, thereby linking nanoelectronic design and fabrication directly to quantum functionality. Predictive device modeling must therefore connect the physical gate architectures to both the electrostatic landscape and the resulting many-electron states. Such a capability is becoming increasingly important for design-technology co-optimization (DTCO) in semiconductor quantum computing, where device geometry, fabrication constraints and quantum functionality must be evaluated within a unified framework[8–11].

Although many recent works have explored the analytical models for symmetric field-effect transistors (FETs), such as double-gate FETs[12] and cylindrical nanowire FETs[13], establishing this connection remains difficult in asymmetric multigate devices. In a typical CMOS qubit chip, gates of unequal width occupy different dielectric interfaces and interact through continuous dielectric interfaces, producing a confinement landscape that cannot generally be reduced to independent gate responses. Numerical electrostatic solvers can represent such structures, but they often obscure how individual geometrical and material parameters shape the potential.[14] Conversely, compact analytical models developed for transistor electrostatics or idealized gate-defined quantum dots rarely enforce charge redistribution across multiple conductors and dielectric interfaces.[15–19] Many-electron calculations then introduce a second disconnect by replacing the device-derived landscape with harmonic or otherwise prescribed confinement.

This separation becomes particularly restrictive for CMOS qubit chips with Jellybean-type silicon quantum couplers, where long-range spin transfer is required[20]. Such structures are being explored as extended qubit couplers and as artificial systems in which interactions, valley physics and confinement compete to determine shell filling.[15,17,21,22] Their spatially non-uniform potentials can generate occupation-dependent spin states and multipeak charge distributions that are absent from conventional circular-dot shell models. A unified framework is therefore needed to trace these features back to the gate geometry, and potentially traps and other defects nearby, rather than treating electrostatics and electron filling as unrelated simulation stages.

Here we develop a reduced-order analytical framework that connects an asymmetric multigate CMOS stack directly to many-electron shell filling. The electrostatic component, the Poisson-kernel coupled-interface Green-function (PK-GF) model, starts from finite-gate Poisson kernels and uses a coupled-interface Green operator with an effective charge expansion to enforce conductor voltages across the dielectric stack. The same gate-derived potential is then passed to an exact-exchange spin-valley calculation. This creates one engineering chain from layout and bias to confinement and spin occupation.

This electrostatic-to-spin workflow is the main advance beyond an electrostatic fitting exercise. We validate the analytical model against an independent finite-volume solution of the same boundary-value problem. We then use the jellybean quantum-dot geometry to test whether the validated confinement supports Wigner-molecule-like charge localization and a low-spin branch relevant to a long-range coupler. Unrestricted Hartree-Fock (UHF) captures the localization but likely overestimates spin polarization. Complete active-space configuration interaction (CASCI) supports low-spin, including S = 0, solutions within the accessible active spaces. It does not prove the correlated spin ground state when the competing high-spin sector lies outside that space.

## Analytical methods for electrostatics of multi-gate CMOS qubit devices

To explicitly present our model, we take a four-gate silicon quantum device based on the jellybean-dot geometry as an example: two plunger gates and two exchange (J) gates. As shown in Fig. 1, the model is defined by an explicit two-dimensional boundary contract. The silicon substrate has thickness H = 2000 nm, with a grounded bottom boundary, and is overlaid by a continuous 6 nm $SiO_2$ layer, a lower 4 nm $Al_2O_3$ dielectric layer, zero-thickness plunger gates G2 and G4 at y = 10 nm, a 4 nm $Al_2O_3$ intergate dielectric layer, and zero-thickness exchange gates G1 and G3 at y = 14 nm, consistent with reported CMOS gate-defined silicon qubit devices. [18,19] The gate intervals are half-open: G1 = [0, 80) nm, G2 = [83, 233) nm, G3 = [236, 276) nm and G4 = [279, 319) nm. The optional drain boundary is represented by a prescribed profile $V_D(x)$; all reported calculations set $V_D(x) = 0$, so the final potential equals $\varphi_{PK-GF}$.

In our model, a single finite gate first contributes the Poisson-kernel (PK) response in the silicon substrate. Linear superposition of the four finite intervals gives a gate-resolved basis before any numerical discretization is introduced. The lower-order PK model propagates each isolated gate spectrum through the exact planar multilayer transfer but sets the uncovered part of the corresponding effective plane to zero. This zero-continuation assumption is the main limitation of PK: it treats a multilayer interface as if only the metal electrode carries electrostatic information, as in simpler patterned-gate electrostatic models.[4,23–26] To mitigate this issue, we introduce the intermediate PK-prior model, which replaces the missing uncovered region with a device geometry-derived background estimate. This prior-knowledge-driven approach is useful for interface background recovery, but it is not a coupled conductor solution and can still leave significant errors.

We therefore introduce PK-GF as the coupled-interface extension of PK. This method treats the uncovered interfaces and electrodes as a coupled boundary system, combining analytical electromagnetics, boundary integral methods and spectral methods to construct the Green operator. In such a system, the homogeneous-layer elimination produces a four-interface Dirichlet-to-Neumann matrix $\mathrm{D(k)}$ and a matrix Green operator $\mathrm{G(k)} = \mathrm{D^{-1}(k)}$. The unknown gate charge density on each metal electrode is expanded in a gate-width-mapped Chebyshev family with the correct inverse-square-root edge behavior. With $\mathrm{M_q} = 12$ modes per gate, the electrode solve only contains 48 unknown amplitudes and enforces the prescribed gate voltages through $\mathrm{Ka} = \mathrm{V}$. The coefficients $\mathrm{a_{jm}}$ are therefore physical charge-expansion coefficients. They are not adjustable voltage corrections.

We also independently tested three approximations based on the Green function: the exact finite-gate model used by PK, the geometry-derived boundary estimate used by PK-prior and the interface coupling plus finite-order charge expansion used by PK-GF. After establishing the final model, PK-GF, we validated our analytical model using the finite-volume method. The finite-volume field remains outside the analytical construction and is used only for validation.

## Interface coupling and effective charge expansion techniques

As the device architecture studied here is not a two-dimensional (2D) single-gate half-plane model, the exact planar transfer and the equivalent-oxide-thickness approximation agree only at low spatial frequency $|\mathrm{k}|$. At higher $|\mathrm{k}|$, the multilayer transfer retains interface reflections and dielectric filtering that cannot be captured by a single exponential length, as shown in Fig. 2a-c. This is why the production calculation keeps the planar multilayer operator rather than collapsing the stack to an effective oxide thickness.

The larger error, however, comes from boundary continuation rather than from the finite-gate kernel itself. The PK approximation imposes the voltage on each gate interval but suppresses the uncovered part of the effective plane, as shown in Fig. 2d. Near $\mathrm{y} = -2$ nm (electron channel's position) this produces a broad underestimation of the interface-supported potential, as shown in Fig. 2g. Fig. 2e suggests that PK-prior recovers much of this missing background by assigning non-zero uncovered-interface potential from geometry-derived influence weights, but it still does not solve the conductor-charge redistribution problem (see Fig. 2h). PK-GF instead represents each gate charge using edge-aware Chebyshev basis functions (Fig. 2f). The symmetric $m = 0$ mode captures common edge accumulation, whereas higher even and odd modes describe center-to-edge and left-to-right charge redistribution. These curves are structural basis functions, not solved charge profiles; their amplitudes are determined simultaneously from the prescribed conductor voltages. The coupled-interface correction is therefore spatially broad rather than confined to gate edges (Fig. 2g). At $y = -2$ nm, PK underestimates the potential because it sets uncovered effective-plane regions to zero, while PK-prior restores much of this background and PK-GF additionally solves the coupled charge redistribution (Fig. 2h).

The modal order was chosen using internal convergence only. As shown in Fig. 2i, increasing $M_q$ from 10 to 12 changes the potential in the region of interest by 0.081 mV root-mean-square (RMS), whereas increasing $M_q$ from 12 to 14 changes it by a further 0.034 mV. The condition number increases from $5.35 \times 10^3$ at $M_q = 12$ to $1.24 \times 10^4$ at $M_q = 14$. Twelve modes per gate therefore provide an engineering knee between conductor enforcement, field convergence and numerical conditioning. At this order, the gate-voltage residual is 1.698 mV RMS and has a maximum absolute error of 17.456 mV. These results identify the physical role of the coupled-interface correction: It restores the non-local electrostatic response of the uncovered dielectric interfaces and redistributes conductor charge consistently across the four-gate stack.

## Independent finite-volume validation

The analytical PK-GF model was tested against a conservative five-point finite-volume (FV) solution of the same idealized two-dimensional boundary-value problem in an in-house technology computer-aided design (TCAD) tool, as depicted in Fig. 3. Both calculations use the same continuous dielectric stack, half-open gate intervals, prescribed gate voltages, grounded bottom boundary of the silicon substrate and outer-boundary conditions. The validation bias is V1 = 0.80 V, V2 = 0.35 V, V3 = 1.05 V and V4 = 0.55 V. The fine FV grid uses dx = dy = 1 nm and contains approximately 6.49 million volume nodes. The FV matrix is assembled and solved before any analytical array is loaded; no analytical field is used as a boundary value, initial guess, parameter-selection target or residual correction.

The PK-GF and FV potential maps agree closely across the validation domain (Fig. 3a, b). Their residual remains at the millivolt scale (Fig. 3c), whereas the PK residual reaches hundreds of millivolts and retains a broad spatial structure (Fig. 3d). Over the registered region $-100 \leq \mathrm{x} \leq 420$ nm and $-80 \leq \mathrm{y} \leq -3$ nm, the potential root-mean-square error (RMSE) decreases from 239.453 mV for PK to 46.017 mV for PK-prior and 1.134 mV for PK-GF. The PK-GF mean and maximum absolute errors are 1.026 and 2.699 mV, respectively. Its RMSE is also below the 1.539 mV RMS difference between the independently calculated 2 nm and 1 nm FV grids.

The line-cut comparisons test whether this agreement persists across different spatial directions and distances from the gate stack. Horizontal profiles at five depths show that PK-GF reproduces the FV potential across gate edges, inter-gate gaps and the central confinement region (Fig. 3e, f). Vertical profiles at five lateral positions confirm the same agreement throughout the silicon substrate (Fig. 3h, i). The aggregate metrics in Fig. 3g, j show that the improvement is distributed across the validation region rather than arising from isolated matching points. Consistent agreement is also obtained for the electric field: the PK-GF field RMSE is $5.68 \times 10^{-6}$ V nm$^{-1}$, compared with $1.46 \times 10^{-3}$ V nm$^{-1}$ for PK (Fig. 3k, l).

The exact-node comparison at $\mathrm{y} = -2$ nm provides the most stringent test near the dielectric stack (Fig. 3m-p). At this depth, the RMSE decreases systematically from 264.797 mV for PK to 62.284 mV for PK-prior and 1.283 mV for PK-GF. The PK-to-PK-prior improvement demonstrates the importance of restoring a non-zero potential on uncovered interfaces. The remaining PK-prior error shows that this background estimate alone is insufficient. Millivolt-scale agreement is reached only when PK-GF solves the coupled redistribution of conductor charge through the shared interface operator.

PK-GF achieves millivolt-scale agreement with the independent numerical solution for the matched geometry, bias and boundary conditions, without FV-based calibration. This is numerical validation of the analytical electrostatic model, not experimental validation. Defects, disorder, mobile charge and three-dimensional fabrication details remain outside the present boundary-value problem.

## Electrostatics-driven exact-exchange filling and correlated spin branch

Elongated silicon quantum dots provide a complex regime in which confinement geometry, Coulomb interactions, spin and valley states jointly determine electron filling[15,17]. We therefore extend the analytical electrostatic framework into an exact-exchange spin-valley model for N = 2-17 electrons at B = 5 T to examine the electrostatic-to-quantum workflow. Rather than introducing a harmonic or fitted confinement potential, the calculation retains the causal chain from the asymmetric CMOS gate stack to the many-electron states. The gate-derived confinement Hamiltonian is diagonalized once for its lowest 45 spatial valley-orbitals, direct-Coulomb and exact Fock-exchange integrals are evaluated with a fast-Fourier-transform (FFT) Poisson solver, and the UHF problem is solved across four spin-valley channels with direct inversion in the iterative subspace (DIIS). The same softened Coulomb kernel is used for Hartree and Fock terms, so one-electron self-interaction cancels exactly.

For this filling regime, G1, G3 and G4 are biased at −2 V, while G2 is biased at 3 V. These voltages generate an elongated potential well beneath G2, bounded by analytical barrier maxima at x = 28 nm and x = 284 nm (Fig. 4a–d). The PK-GF potential at y = −2 nm is converted into electron confinement energy through

$$U_x(x) = -\alpha_{\text{lever}}\phi_{PK-GF}(x, y_0)$$

where y0 = -2 nm and the lever arm is set to 0.052 $eV\cdot V^{-1}$, consistent with the reported experiment[15]. A smooth transverse aperture preserves this analytical profile along z = 0 and raises the energy toward the barrier ceiling at z = ±20 nm (Fig. 4e). The resulting potential retains the asymmetric gate-defined longitudinal structure while providing finite transverse confinement for the many-electron calculation.

The compact UHF equation used for the Fig. 4 filling calculation is

$$\widehat{\mathrm{H}}_s[\mathrm{n}] = \widehat{\mathrm{H}}_{\mathrm{kin}} + \mathrm{U}_{\mathrm{PK-GF}}(\mathrm{x}, \mathrm{z}) + \widehat{\mathrm{H}}_{\mathrm{valley}} + \mathrm{V}_{\mathrm{H}}[\mathrm{n}] + \widehat{\mathrm{V}}_{\mathrm{x}}^{(s)} - \frac{\mathrm{s}}{2}\mathrm{g}\mu_{\mathrm{B}}\mathrm{B}, \qquad \mathrm{s} = \pm 1$$

Here the Hartree term is built from the total density, while the exchange term is a nonlocal exact Fock operator restricted to the same spin and valley channel. The full operator construction and validation are given in the Supporting Information. At 5 T and g = 2, the Zeeman splitting is 0.579 meV.

Interactions substantially reorganize the shell-filling sequence (Fig. 4f, g, j). The lowest-energy UHF sequence over N=2-17 gives $S_z$ = 1, 1.5, 1, 0.5, 2, 3.5, 4, 4.5, 4, 5.5, 5, 6.5, 7, 7.5, 7 and 8.5. This strong and irregular polarization is not imposed by the noninteracting reference; it emerges from the self-consistent Coulomb and exact-exchange problem. However, the same B=5 T states audited with frozen-core CASCI at the canonical 12-orbital active-space target show a near-minimal correlated spin, $S_z$ between 0 and 1 across N=2-17. At this active-space size, CASCI yields a triplet instead of a singlet only for N = 2, 6 and 10 in the Fig. 4 overlay. Figure 5 examines this discrepancy.

Most importantly, the Hartree energy, built from the total density, rises from 0.019 eV at N=2 to 0.806 eV at N=17, whereas the exact-exchange energy, evaluated on the broken-symmetry UHF determinant, grows in magnitude from -9.7 meV to -192 meV and thus tracks the overestimated UHF spin polarization rather than a converged exchange (Fig. 4h). Over the same range, the longitudinal density spread increases from 15.5 nm to 33.0 nm, and the transverse spread from 6.3 nm to 8.4 nm (Fig. 4i). The mean position remains near the center of the well, showing that the density expands mainly along the elongated confinement rather than along the transverse z direction. Occupation-dependent longitudinal maxima emerge within an extended, gate-aligned density envelope (Fig. 4f, k). The robust UHF result is the spatial localization, not the spin assignment, which is tested by UHF and CASCI in Fig. 5. The classical localization proxy in Fig. 4l provides a strong-coupling comparator for the multi-center artificial molecule reported experimentally[15].

In the UHF calculation, fifteen of the sixteen UHF occupations satisfy all stored convergence criteria; the fully spin-polarized N=15 case settles into a benign self-consistent limit cycle with a final energy change of 0.067 meV and is reported at its lowest-energy iterate. Each N is an independent calculation.

The results therefore establish an integrated route from asymmetric CMOS gate geometry-based electrostatics to interaction-dependent shell filling and jellybean-like localization tendencies. They do not certify an experimentally realized Wigner molecule or a uniquely converged correlated spin ground state; those stronger claims require orbital magnetic coupling, enlarged active spaces, spin/space symmetry restoration, multi-seed stability analysis and comparison with measured charge and spin transitions.

## Correlated spin filling modeling

Long-range coupling is the second device function sought for the jellybean quantum dot, and the analysis in Fig. 4 has already established its spatial prerequisite: the gate-derived well stays continuous between the two ends, while added electrons expand mainly along $\mathbf{x}$ and build multiple longitudinal maxima within a single gate-aligned density envelope (Fig. 4a-e, i, k, l).

What this leaves open is whether the extended density acts as one correlated mediator and whether its spin is quiet enough to relay a qubit state. Figure 5 tests both points by applying two complementary treatments to the same gate-derived states: the exact-exchange UHF of Fig. 4, now read for its localization and spin structure (Fig. 5a-l), and a frozen-core active-space CASCI that treats near-Fermi correlation within the selected active space and tests the mean-field spin assignment (Fig. 5m-r). The two treatments support the same charge-localization picture but, as shown below, disagree on the spin, and that contrast is the specific contribution of Fig. 5 over Fig. 4.

The localization diagnostics again identify a continuous interacting system rather than a set of disconnected dots (Fig. 5a-f). The longitudinal density carries up to eight resolved maxima at $N = 17$, and from $N = 2$ to 17 the lobe count grows while the interaction-to-level-spacing ratio $R = E_C/\Delta$ stays between 2.5 and 4.4. The conditional correlation hole, the modulated addition spectrum, and a broken-symmetry energy gain reaching about 245 meV independently place this elongated density in an interaction-dominated, Wigner-molecule-like regime. The lobe count occasionally drops, for instance past its $N = 14$ maximum, when the two-row density reorganizes into a zigzag pattern and neighboring maxima merge in the one-dimensional projection; these reversals are internal rearrangements of the continuous envelope, not a loss of mediation.

The same UHF solution fixes the geometry of the mediator but not its spin. At the physical silicon interaction strength, the UHF ground state is polarized, and the constrained $S = 0$ branch is more extended and carries fewer lobes yet for $N = 12$ sits only 23.9 meV above it (Fig. 5g-i, l). Two scans expose the origin of this ordering. In the interaction scan all tested fillings are spin quiet at $\kappa = 0.25$, polarization appears by $\kappa = 0.5$, and all three fillings are polarized by $\kappa = 0.75$ (Fig. 5j), so the polarization is a Stoner-type response to interaction strength rather than a structural feature. Varying the valley splitting from 0.12 to 8 meV produces no systematic suppression (Fig. 5k), which indicates that valley splitting is not the controlling factor over this range. The re-entrant points at larger $\kappa$ and the crossings in the valley scan follow switches between independently optimized determinants, not genuine operating windows. UHF therefore supports the localized mediator but likely overestimates its polarization, which is what the correlated treatment is needed to test.

Active-space CASCI provides a correlated check and favors a low-spin assignment (Fig. 5m-o). In the no-frozen-core test, where the fully polarized competitor lies inside the active space, the calculated ground state is an $S = 0$ singlet for $N = 2,4,6$ and 8, with the polarized state higher by 27.3, 25.4 and 34.7 meV for $N = 4,6$ and 8. The 0.17 meV separation at $N = 2$ is an isolated near-degeneracy that leaves only that endpoint sensitive to omitted perturbations and does not overturn the low-spin trend. The comparison stays fair at $N = 12$ for $n_{act} = 12$, where the cap $S_{max} = 6$ still contains the UHF $S_z = 5$ competitor and the accessible high-spin sector lies 142.8 meV above the singlet. For $N = 16$ and 20 the full UHF high-spin state falls outside the active space, so those bars probe only the accessible high-spin sector and not the polarized competitor itself.

Sweeping the active space separates what is converged from what is not (Fig. 5p-r). The ground state remains $S = 0$ for every tested $N$ and $n_{act}$, and the high-spin gap stays positive throughout, generally above 150 meV at $n_{act} = 12$; this gap is not a strict variational bound, because the frozen core shifts with the active space, but its sign and scale are stable. The singlet-triplet gap behaves differently, swinging non-monotonically from 0.27 to 33.3 meV, with the 33.3 meV value at $N = 12$ and the 0.27, 10.1, 7.9 and 15.7 meV sequence at $N = 16$ reflecting uneven convergence of the differential correlation energy between spin sectors, but the trends are similar. The low-spin ordering is robust across the tested active spaces, whereas the quantitative singlet-triplet splitting is not converged.

Fig. 5 supports a spin-quiet correlated branch for the extended mediator. It does not establish an operational long-range coupler or prove the ground-state spin against competitors excluded by the active-space cap. The present calculation also does not provide the two end-qubit exchange couplings, their electrical on/off ratio, disorder tolerance or a converged finite-field excitation gap.

The active-space accessibility boundary used to delimit the correlated claim is

$$S_{max} = \frac{1}{2}\min(n_e, 2n_{act} - n_e), \qquad n_{act} = 12 \text{ for the largest reported active space.}$$

where $n_e$ is the active-electron count and $n_{act}$ is the number of active spatial-valley orbitals. Thus, CASCI can exclude a UHF high-spin competitor only when that competitor lies within this bound; otherwise, the correlated low-spin branch is supported but not proven against the inaccessible state.

## Discussion

A gate-resolved path from process geometry to quantum-device observables is hereby established. A finite-gate Poisson kernel captures electrode extent, but it misses electrostatic coupling through the uncovered dielectric interfaces. PK-prior restores much of that background. PK-GF also solves conductor-charge redistribution across the shared interface system. This final step reduces the potential RMSE from 239.453 mV for PK and 46.017 mV for PK-prior to 1.134 mV against the matched FV calculation. No FV field or residual enters the analytical construction.

That distinction matters for DTCO. Every input has a direct device meaning: gate position and width, dielectric material and thickness, applied voltage, or the prescribed drain boundary. The operator-level solution remains analytical, and the numerical work is limited to spectral evaluation and a 48-coefficient conductor solve. The model can therefore expose which layout or process choice changed the confinement, rather than returning only a field map. It is also structured for geometry and bias sweeps, although this study does not report a timing benchmark.

The second contribution is the handoff from device electrostatics to many-electron physics without replacing the gate-defined profile by harmonic confinement. In the examined jellybean quantum dot, several diagnostics support an occupation-dependent, Wigner-molecule-like charge density with multiple longitudinal maxima. That localization persists across the mean-field and correlated analyses. The spin assignment does not. UHF likely overestimates polarization, while CASCI supports a low-spin branch within the tested active spaces. The quantitative singlet-triplet gap remains unconverged.

This separation between robust and method-sensitive outputs is useful for quantum-device DTCO. Charge localization can be treated as a design signal in the present model, whereas UHF spin polarization cannot. A practical workflow should use PK-GF to screen how layout, dielectric stack and bias reshape confinement, then apply a correlated calculation before using spin as an optimization target. The model therefore connects process choices to quantum objectives without hiding the point at which a higher-fidelity calculation is needed.

The scope remains limited. The electrostatic model is two-dimensional, linear and free of mobile charge in the silicon substrate and dielectric layers. It excludes quantum capacitance, interface traps, disorder, finite metal thickness, process variability and three-dimensional gate-end fields. The many-electron calculation inherits these limits. It also treats the magnetic field through the Zeeman term only, truncates the orbital basis and cannot represent every high-spin competitor at large electron number. The correlated result therefore supports, but does not prove, the low-spin branch outside the fair active-space tests.

## Conclusion

By linking multigate CMOS geometry to independently validated electrostatics and then to correlation-audited many-electron observables, this work provides a reduced-order bridge between device design and qubit function. PK-GF reproduces the matched two-dimensional FV solution at millivolt accuracy without calibration to that field. The downstream analysis then combines charge localization with spin assignments that is the key to understand the spin features of the qubit devices. By doing so, we demonstrate a practical DTCO contribution: layout and process variables can be screened against physically meaningful confinement and coupler targets before full three-dimensional simulation and experiment. But, it should be admitted that the framework is not yet a complete DTCO flow which requires self-consistent carrier electrostatics, fabricated three-dimensional geometry, disorder and variability analysis, end-qubit exchange metrics and experimental tests of the correlated spin states.

## Methods

### Device and boundary contract

The silicon substrate occupies $-H \le y \le 0$ with H = 2000 nm and relative permittivity $\epsilon_{\mathrm{Si}} = 11.7$. The bottom boundary of the silicon substrate is grounded. Above the silicon substrate are a continuous 6 nm $SiO_2$ layer with $\epsilon_{\mathrm{SiO_2}} = 3.9$, a continuous 4 nm $Al_2O_3$ layer with $\epsilon_{\mathrm{Al_2O_3}} = 9.0$, zero-thickness plunger gates $G_2$/$G_4$ at y = 10 nm, a continuous 4 nm $Al_2O_3$ intergate dielectric layer, zero-thickness exchange gates $G_1$/$G_3$ at y = 14 nm, and air to $y_{\mathrm{top}} = 400$ nm. Top and lateral outer boundaries use zero normal flux.

The half-open intervals are $G_1$ = [0, 80) nm, $G_2$ = [83, 233) nm, $G_3$ = [236, 276) nm and $G_4$ = [279, 319) nm. The validation voltages are 0.80, 0.35, 1.05 and 0.55 V. The general boundary contract includes a prescribed drain boundary $\phi(x, y_D) = V_D(x)$ on $\Omega_D$, but all reported figures set $V_D(x) = 0$.

### Poisson-kernel and prior models

For $y < 0$, the finite-interval response of gate $j$ with endpoints $L_j$ and $R_j$ is

$$A_j(x,y) = \frac{1}{\pi}\left[\tan^{-1}\left(\frac{R_j - x}{-y}\right) - \tan^{-1}\left(\frac{L_j - x}{-y}\right)\right].$$

The finite silicon thickness is enforced by the grounded-substrate propagator

$$G_H(k,y) = \frac{\sinh[|k|(y+H)]}{\sinh(|k|H)}, \qquad G_H(0,y) = \frac{y+H}{H}.$$

The independent Poisson-kernel model is

$$\phi_{\mathrm{PK}}(x,y) = \mathcal{F}^{-1}\left\{G_H(k,y)\sum_{j=1}^{4} V_j\,\hat{\chi}_j(k)T_j(k)\right\}.$$

Its defining approximation is zero potential outside each gate interval on the independently propagated effective plane. The prior model replaces that zero continuation with a geometry-derived background. The mean equivalent oxide thickness (EOT) appears in the background weight:

$$u_{\text{prior}}(x) = \frac{\sum_{j=1}^{4} V_j\, w_j(x)}{\sum_{j=1}^{4} w_j\,(x) + w_{\text{bg}}}, \qquad w_{\text{bg}} = \frac{\bar{t}_{\text{EOT}}}{H} = 0.0129.$$

The prior uses only geometry and material permittivity and is not passed to the PK-GF solve.

### Coupled-interface Green operator and conductor solve

For interface potentials $\mathbf{u}(k) = [u_0, u_6, u_{10}, u_{14}]^{\mathrm{T}}$, homogeneous-layer elimination gives

$$\mathbf{q}(k) = \mathbf{D}(k)\mathbf{u}(k), \qquad \mathbf{G}(k) = \mathbf{D}^{-1}(k).$$

The entries of $\mathbf{D}(k)$ are formed from $A_\ell(k) = \epsilon_\ell |k| \coth(|k| d_\ell)$, $B_\ell(k) = \epsilon_\ell |k| \operatorname{csch}(|k| d_\ell)$, the grounded-substrate load and the top-air load. On gate $j$, the effective conductor charge is expanded as

$$\sigma_j(x) \approx \sigma_j^{\text{eff},(12)}(x) = \frac{\chi_j(x)}{\sqrt{1 - \xi_j^2(x)}} \sum_{m=0}^{11} a_{jm}\, T_m\big[\xi_j(x)\big], \qquad \xi_j(x) = \frac{2\big(x - x_{c,j}\big)}{W_j}.$$

Applying the Green operator to each basis function and collocating the prescribed conductor voltages gives

$$K\mathrm{a} = \mathrm{V}.$$

The system is solved by singular-value-decomposition least squares. No FV field, residual, coordinate shift, fitted voltage, fitted geometry or fitted material parameter enters $K$, a or V.

### Final electrostatic representation

The prescribed drain contribution is

$$\phi_D(x, y; V_D) = \frac{1}{\pi} \int_{\Omega_D} \frac{-(y - y_D) V_D(x')}{(x - x')^2 + (y - y_D)^2}\, dx'.$$

The complete representation is

$$\phi_{\text{final}}(x, y; V_D) = \mathcal{F}^{-1} \left\{ G_H(k, y) \mathrm{e}_0^{\mathrm{T}} \mathrm{D}^{-1}(k) \sum_{j=1}^{4} \sum_{m=0}^{11} a_{jm}\, \hat{\sigma}_{jm}(k) \mathrm{e}_{\tau(j)} \right\} + \phi_D(x, y; V_D).$$

For every reported figure,

$$V_D(x) = 0 \quad \Rightarrow \quad \phi_{\text{final}}(x, y) = \phi_{\text{PK-GF}}(x, y).$$

## Independent FV validation

The FV solver independently discretizes $\nabla \cdot [\epsilon(x, y)\nabla\phi] = 0$ with a conservative five-point scheme. The fine grid uses 1 nm spacing and the coarse grid uses 2 nm spacing. Analytical and numerical fields are compared on exact common nodes using the same half-open gate convention; interpolation and nearest-node substitution are not used. Metrics are evaluated over $-100 \leq x \leq 420$ nm and $-80 \leq y \leq -3$ nm, with an additional exact cut at $y = -2$ nm.

## Many-electron application

For the many-electron demonstration, $G_1/G_3/G_4 = -2$ V and $G_2 = 3$ V. The PK-GF profile at $y = -2$ nm is converted to electron energy as

$$U_x(x) = -\alpha_{\text{lever}}\phi_{PK-GF}(x, -2\text{ nm}), \qquad \alpha_{\text{lever}} = 0.052\text{ eV} \cdot \text{V}^{-1}$$

The computational interval is bounded by the analytical barrier peaks at x=28 nm and x=284 nm. The two-valley effective-mass calculation uses dx=0.4 nm, effective dy=0.8 nm, dz=0.2 nm, B=5 T, g=2, valley splitting 0.12 meV and valley coupling 0.06 meV. The spatial Hamiltonian is diagonalized for its lowest 45 orbitals by Chebyshev-filtered subspace iteration. Direct-Coulomb and exact same-spin, same-valley Fock-exchange integrals are evaluated once by FFT-Poisson, and the UHF equations are solved in this orbital basis with DIIS. Figure 4 overlays the same-field frozen-core CASCI spin computed with the 12-orbital active space. Figure 5 reports the B=0 active-space convergence, no-frozen-core fair test and high-spin accessibility caveats. UHF convergence requires energy, per-electron density and exchange changes below $10^{-4}$, $10^{-4}$ and $10^{-3}$ meV; the selected N=15 iterate is reported with the stated limit-cycle caveat.

## Acknowledgement

We used large language model such as ChatGPT (OpenAI) for English language editing and proofreading. All scientific content, analysis, interpretations, and conclusions were developed, verified, and approved by the authors.

**Figure 1 | Asymmetric multi-gate CMOS qubit device and the formula-first PK-GF workflow.** The device schematic fixes the boundary contract: a grounded silicon substrate, SiO2 and Al2O3 dielectric layers, plunger (P) and junction (J) gates on two interface planes, the half-open gate intervals, and the prescribed drain boundary that represents an electron reservoir. The potential is assembled analytically, beginning from the finite-gate Poisson-kernel (PK) response and its grounded-Si substrate propagator to form a multi-gate basis. Two reference reductions, PK and PK-prior, isolate how the treatment of the uncovered interfaces and the conductor charge controls accuracy. The production model, PK-GF, adds a four-interface Dirichlet-to-Neumann mapping, a matrix coupled-interface Green-function operator and an edge-aware Chebyshev effective-charge expansion, yielding the closed PK-GF master equation. The finite 12-mode charge expansion is the only production approximation, and no finite-volume field or residual enters its construction. A spin filling modeling adopts the electrostatics from the PK-GF and demonstrate an end-to-end complete analytical modeling framework for CMOS qubit devices.

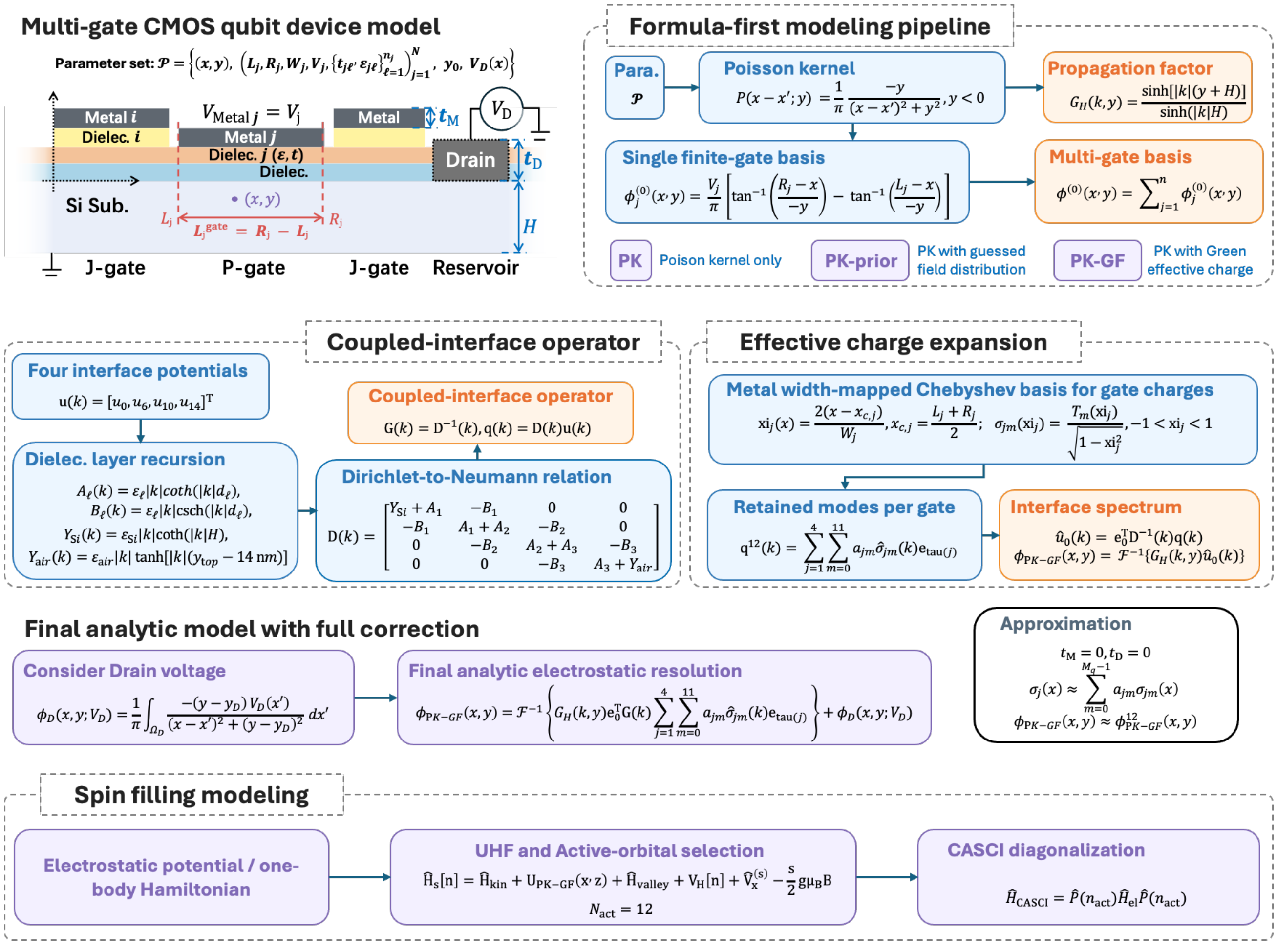

**Figure 2 | Construction of the PK, PK-prior and PK-GF models. a,** Device cross-section. **b,** Finite-gate Poisson-kernel basis responses for each individual gate electrode. **c,** Exact multilayer planar transfer and its equivalent-oxide-thickness limit. **d,** PK independent-gate boundary approximation, which sets the uncovered surface to zero. **e,** PK-prior estimate of the uncovered surface from geometry alone. **f,** Edge-aware Chebyshev functions used in the PK-GF effective-charge expansion. **g,** Coupled-interface correction, the difference between PK-GF and PK. **h,** Progression of the three analytical potentials along the qubit plane at y=-2 nm. **i,** Internal conductor residual, successive field change and operator conditioning used to select 12 modes per gate; the finite-volume error plays no part in this selection.

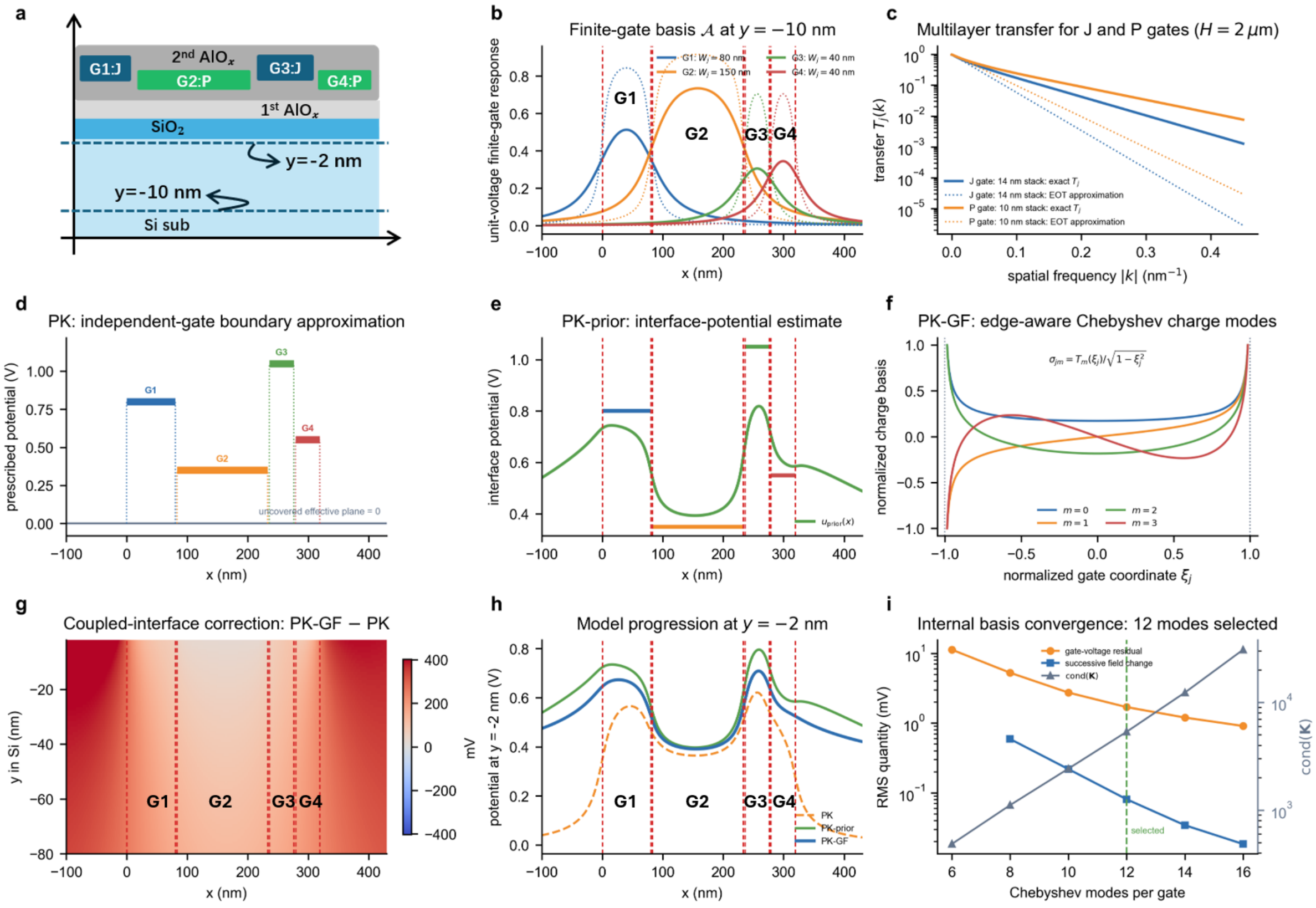

**Figure 3 | Independent finite-volume validation of PK-GF. a-d,** PK-GF and FV potential maps with their residual over the registered domain. **e-j,** Five-depth horizontal and five-position vertical line cuts with aggregate error metrics. **k,l,** FV electric-field magnitude and depth-resolved field RMSE. **m-p,** Exact y=-2 nm comparison of PK, PK-prior, PK-GF and FV, with signed and absolute errors and cut-line RMSE. The analytical and FV grids are exactly co-registered, and FV data enter neither the analytical construction nor the mode-order selection; agreement reaches the millivolt scale without any prior fit to the FV solution.

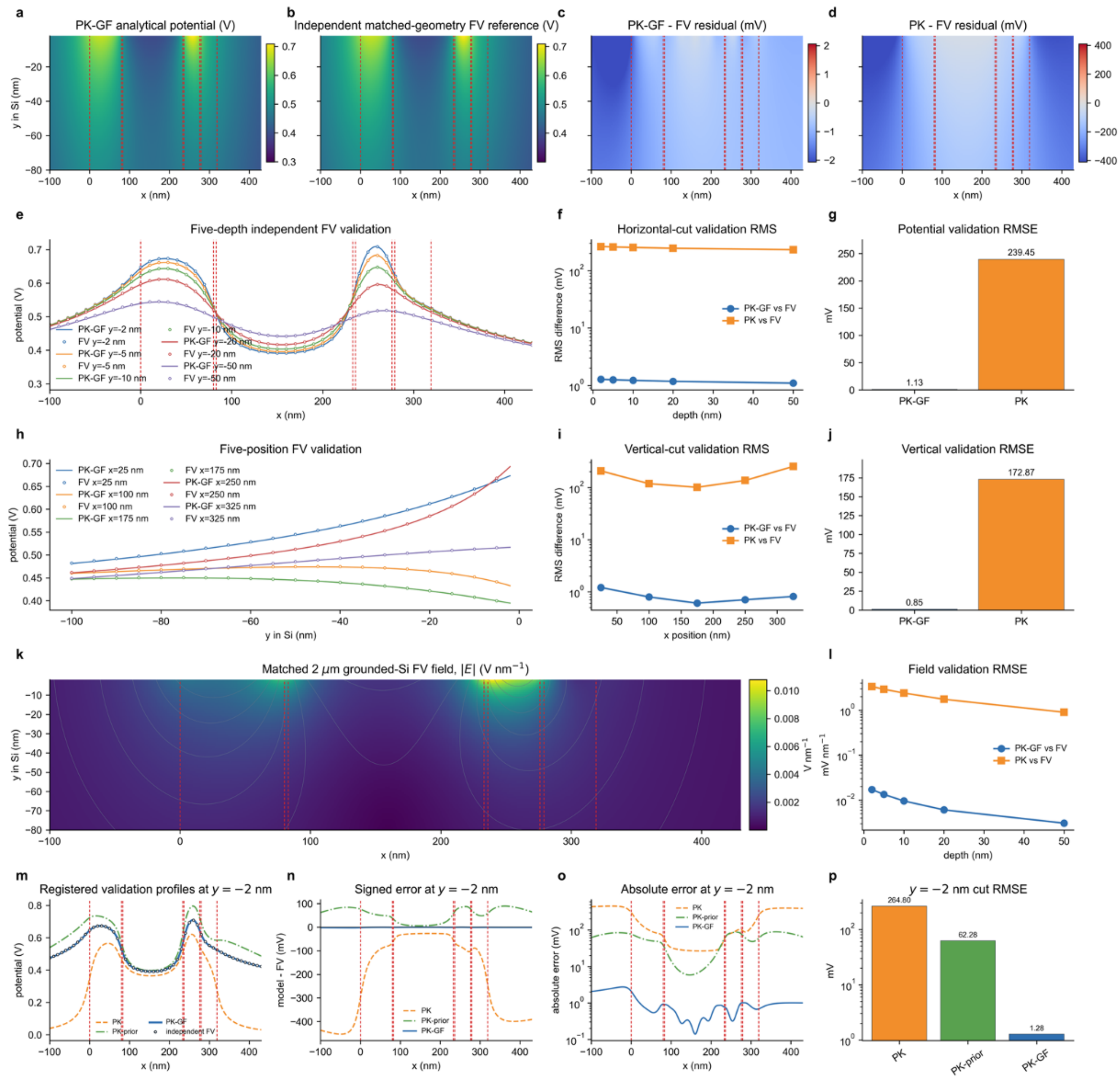

**Figure 4 | Electrostatics-driven exact-exchange filling at B=5 T. a-e,** Device structure and gate-derived potential, electric field, qubit-plane line profile and the z-confined potential well for G1/G3/G4=-2 V and G2=3 V. **f,** Highest-occupied spin-density maps across the filling sequence N=2-17. **g,** Noninteracting shell-filling reference. **h,** Hartree and single-particle level-gap energies together with the spin projection Sz; the UHF Sz and the correlated CASCI Sz at the same field are overlaid for direct comparison. **i,** Longitudinal and transverse density extent (UHF). **j,** Spin-sector filling, again overlaying the UHF and CASCI Sz. **k,l,** Representative total electron density and the classical localization proxy. The UHF and CASCI have discrepancy in ground spin state, but all infer that Jellybean qubit can form a continuous channel for spin transfer.

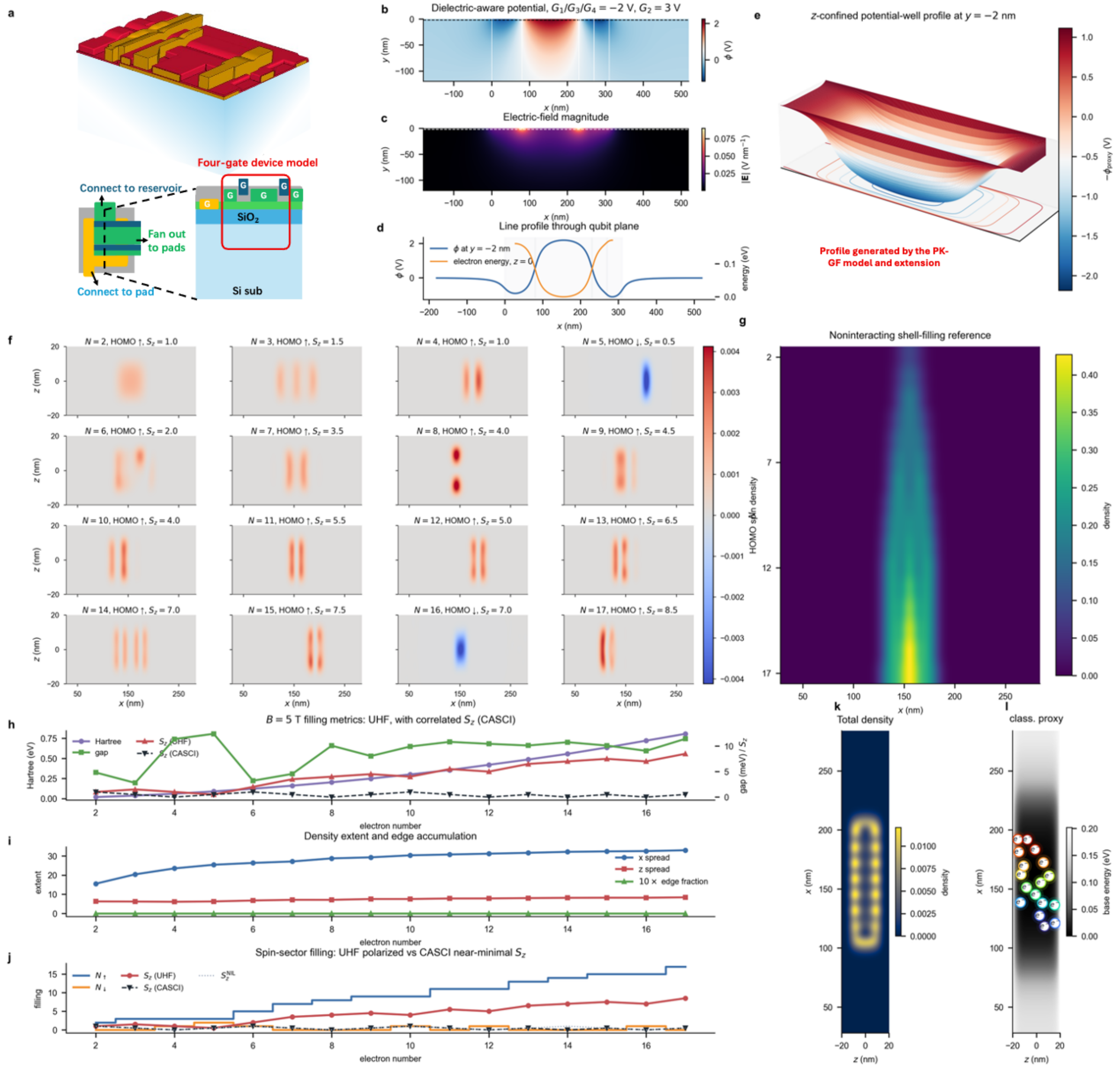

**Figure 5 | Wigner-molecule localization and the correlated spin of the Jellybean mediator.** We combine UHF and CASCI to the same gate-derived states: exact-exchange UHF (**a-l**) and a frozen-core active-space CASCI audit (**m-r**). **a,** Longitudinal charge-density lobe profile. **b,** Resolved lobe count versus electron number N. **c,** Conditional pair correlation n(x|x0) showing the correlation hole that infer the strong interaction. **d,** Interaction-to-level-spacing ratio $R = E_C/\Delta$. **e,** Addition-energy spectrum. **f,** Spin contamination and broken-symmetry energy gain. Panels a-f place the elongated density in an interaction-dominated, Wigner-molecule-like regime. **g, h,** Two-dimensional charge density of the polarized state versus the spin-quiet S=0 singlet at representative N=12, contrasting the chain-like and extended forms. **i,** UHF over-polarization energy and singlet-triplet gap versus N. **j,** Stoner transition of the spin polarization with interaction scale kappa. **k,** Polarization versus valley splitting, which is not the controlling parameter. **l,** Charge-lobe count for the polarized chain versus the extended singlet. **m-o,** Active-space CASCI resolution at the largest active space $n_{act}$=12 (B=0): **m,** corrected ground-state spin versus N, comparing the polarized UHF Sz with the CASCI S=0 result and the no-frozen-core fair test (shaded, N<=8); **n,** energy ladder at $n_{act}$=12, showing the lowest triplet and the accessible high-spin sector above the S=0 singlet for N=12, 16 and 20; **o,** high-spin gap versus N for $n_{act}$=6, 8, 10 and 12. **p-r,** Active-space convergence: **p,** singlet-triplet gap versus $n_{act}$, which is not converged (about 0.3-33 meV); **q,** high-spin gap versus $n_{act}$, robustly large and positive; **r,** ground-state spin versus $n_{act}$, remaining S=0 at every active space and filling. Here $n_e$ is the active-electron count, $n_{act}$ the number of active spatial-valley orbitals.

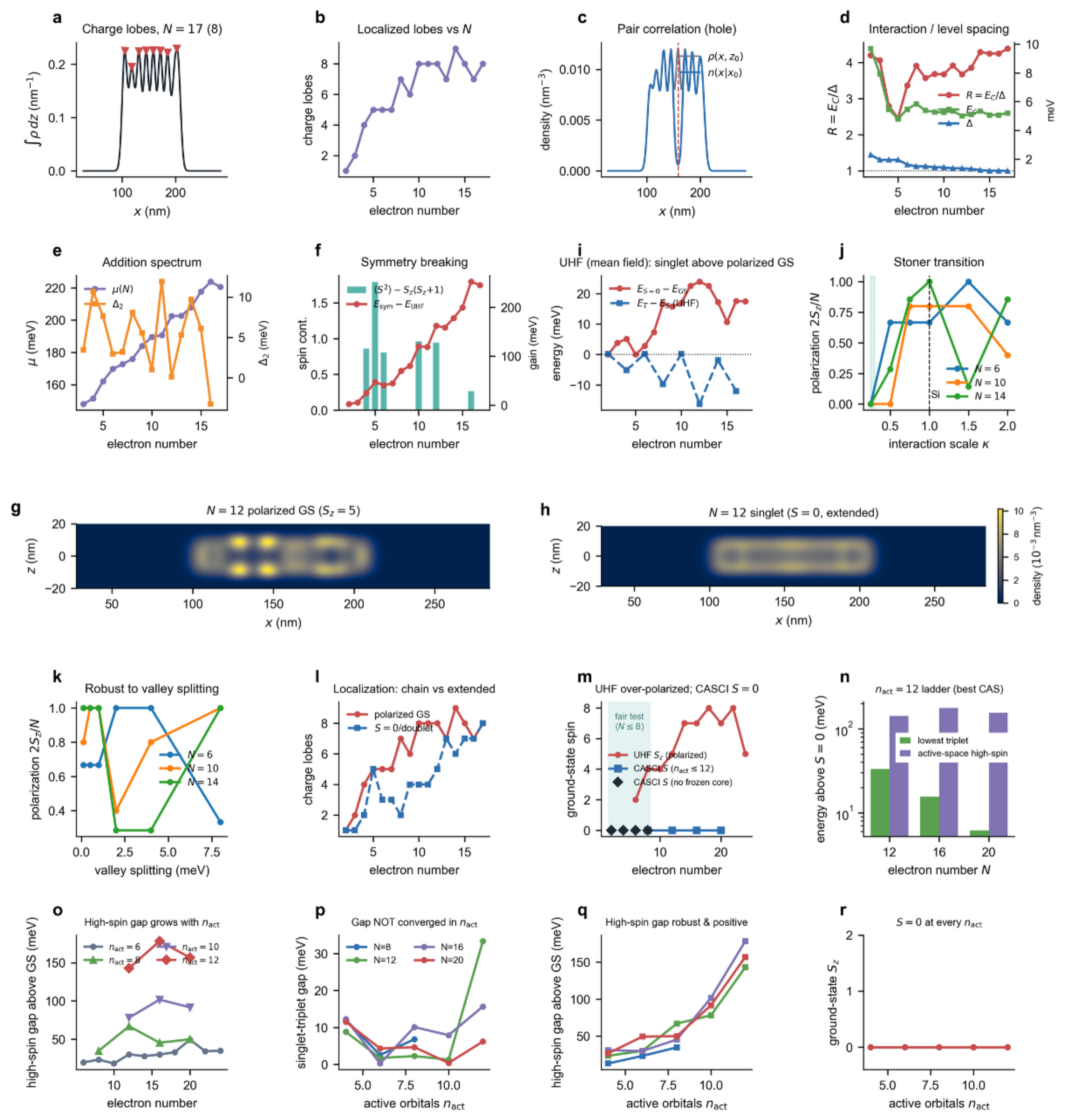
a Charge lobes, $N = 17$ (8)
b Localized lobes vs $N$
c Pair correlation (hole)
d Interaction / level spacing
e Addition spectrum
f Symmetry breaking
i UHF (mean field): singlet above polarized GS
j Stoner transition
g $N = 12$ polarized GS ($S_z = 5$)
h $N = 12$ singlet ($S = 0$, extended)
k Robust to valley splitting
l Localization: chain vs extended
m UHF over-polarized; CASCI $S = 0$
n $n_{act} = 12$ ladder (best CAS)
o High-spin gap grows with $n_{act}$
p Gap NOT converged in $n_{act}$
q High-spin gap robust & positive
r $S = 0$ at every $n_{act}$
electron number
interaction scale $\kappa$
valley splitting (meV)
active orbitals $n_{act}$
x (nm)
z (nm)
density ($10^{-3}$ nm$^{-3}$)